\documentclass[]{icas2022}

\usepackage{caption,booktabs}

\def\BibTeX{{\rm B\kern-.05em{\sc i\kern-.025em b}\kern-.08em
    T\kern-.1667em\lower.7ex\hbox{E}\kern-.125emX}}

\TitlePaper{On the influence of gradient reconstruction procedures over the accuracy of finite volume based schemes}
\AuthorPaper[1]{Frederico Bolsoni Oliveira}
\AuthorPaper[2]{João Luiz F. Azevedo}
\affil[1]{Instituto Tecnológico de Aeronáutica, DCTA/ITA, Pr. Mal. Eduardo Gomes, São José dos Campos, 12228-900, SP, Brazil}
\affil[2]{Instituto de Aeronáutica e Espaço, DCTA/IAE, Pr. Mal. Eduardo Gomes, São José dos Campos, 12228-904, SP, Brazil}

\abstractEnglish{

In the context of the cell centered finite volume approach, care must be taken when performing the reconstruction of property gradients at cell interfaces. The present work analyzes three different gradient reconstruction procedures, using three different turbulent simulation test cases, namely the zero-gradient flat plate, the subsonic NACA 0012 airfoil and the transonic OAT15A airfoil. The analysis is concerned mainly with the usage of quadrilateral meshes. The gas dynamics equations are solved using an implicit implementation of Roe's second-order upwind scheme. The RANS closure problem is solved by using the negative Spalart-Allmaras turbulence model. The solution quality of each gradient discretization procedure is analyzed and compared to experimental data and other numerical solutions available in the literature. For the cases considered here, excellent agreement is obtained between the computed solutions and the expected results, regardless of which gradient reconstruction scheme is used.

}

\begin{document}

%%%%%%%%%% Body of Document%%%%%%%%%%%%%%%%%%%%%%%%
\body 

\section{Introduction}

With the advent of the Industry 4.0, the demand for high-fidelity numerical simulations has been steadily increasing over the past few years among all fields of application \cite{digitalTwinAIAA2020}. Therefore, any modification to well established numerical methodologies that yields improvements over simulation results are of great interest to the industry. In the realm of computational fluid dynamics (CFD), an approach that has been greatly used throughout the decades and that has been demonstrated to be capable of achieving great results is the finite volume method (FV). In the aerospace industry, FV is commonly used to solve systems of conservation laws, among others, the compressible Navier-Stokes equations.

Multiple numerical schemes have been developed over the years in the context of the FV approach \cite{hirschV2}. Those that rely on a cell-centered formulation, though, have one particular trait in common: they require the evaluation of flow properties, and also flow property gradients, at discrete cell interfaces, where those values are not readily available. On a cell-centered formulation, the known discrete properties are taken to be volumetric averages inside each cell. Hence, a reconstruction procedure must be used in order to define suitable values for the unknown properties at each cell interface. Depending on the property in question, the definition of a reconstruction procedure is not necessarily straightforward. This is especially troublesome when the reconstruction of property gradients are considered. On the subject of fluid dynamics, some of the properties that fall into this category are, for instance, the gradients of the velocity components, required in the calculation of the viscous forces. Furthermore, if compressibility effects are taken into account, then the gradient of the fluid internal energy must also be reconstructed during the evaluation of Fourier's law in the energy conservation equation.

It has been previously reported in the literature that the use of different gradient reconstruction techniques can drastically change the outcome of viscous fluid simulations \cite{jalali2014}. The effects vary from changing the overall robustness of the CFD algorithm being employed, to modifying or dissipating fluid structures that are present in the solution field. Unfortunately, though, no single gradient reconstruction procedure has been found so far to be well suited for all situations. The present study is inserted exactly in this context, and aims to provide numerical data for better understanding the effects of different gradient reconstruction techniques over the solution of compressible turbulent flows when applied to quadrilateral meshes. Thus, CFD users can make a more informative decision regarding what gradient reconstruction procedure to use for a given problem configuration.

In the current work, the three-dimensional flow dynamics are modeled using the compressible Reynolds-averaged Navier-Stokes (RANS) equations \cite{hirschV2}.
These equations are discretized in a cell-centered FV framework by using Roe's flux-difference splitting scheme for the reconstruction of the convective fluxes \cite{roe1981, bigarella2009, bigarella2012}. A second-order, total-variation diminishing (TVD), version of the scheme is implemented by using a piece-wise linear reconstruction of the solution \cite{barth1989}, coupled with Venkatakrishnan's limiter \cite{venkatakrishnan1995}. In order to solve the closure problem, inherent to the RANS equations, the negative Spallart-Allmaras (SA-neg) turbulence model is employed \cite{spalart1992, spalart2012, TMR}. 
Discrete cell gradients are computed using a volume-weighted Green-Gauss approach. Property gradients at cell interfaces, however, are computed using three different reconstruction procedures: $A00$, $A0E$ and $AJ0$, whose naming conventions follow Ref.~\cite{jalali2014}, and that will be described in the forthcoming sections.

Simulations are performed using an in-house code, BRU3D \cite{bigarella2009, bigarella2012}, for three different cases. The first one is a two-dimensional turbulent flat plate \cite{TMR}, which is mainly used as a sanity test case. 
The second one is the subsonic NACA 0012 airfoil with a $15$ deg.~angle of attack \cite{TMR}. Its main purpose is to observe the influence that the different procedures have over the value of the aerodynamic coefficients when a separation bubble is present in the solution. The final case is the transonic OAT15A airfoil \cite{roddle1994, bigarella2009, bigarella2012}. It illustrates the performance of each scheme when a shock wave is present in the domain.

This introduction section is followed by a presentation of the numerical formulation used in the present work. Then, a brief description of the test cases is made, accompanied by the obtained results the concluding remarks.

\section{Numerical Formulation}

\subsection{General Formulation of the Method}

The system of conservation laws used here, which for now on will be referred to as the RANS equations, can be written as

\begin{equation} \label{Eq:RANS}
\frac{\partial \vec{Q}}{\partial t} +  \vec{\nabla} \cdot \vec{ \mathcal{F} } ( \vec{Q} , \overrightarrow{\nabla Q} ) \equiv \frac{\partial \vec{Q}}{\partial t} +  \vec{\nabla} \cdot \left[ \vec{\mathcal{F}}_e ( \vec{Q} ) - \vec{\mathcal{F}}_v ( \vec{Q} , \overrightarrow{\nabla Q} ) \right] = 0 \mbox{ ,}
\end{equation}

\noindent
where $\vec{Q}$ is the vector of conserved variables. Furthermore, $\vec{\mathcal{F}}$ is a geometric vector of algebraic vectors, such that

\begin{equation}
\vec{\mathcal{F}}_e ( \vec{Q}) \equiv \vec{E}_e ( \vec{Q} ) \: \hat{i} + \vec{F}_e ( \vec{Q} ) \: \hat{j} + \vec{G}_e ( \vec{Q} ) \: \hat{k} \mbox{ ,}
\end{equation}

\noindent
and

\begin{equation}
\vec{\mathcal{F}}_v ( \vec{Q} , \overrightarrow{\nabla Q} ) \equiv \vec{E}_v ( \vec{Q} , \overrightarrow{\nabla Q} ) \: \hat{i} + \vec{F}_v ( \vec{Q} , \overrightarrow{\nabla Q} ) \: \hat{j} + \vec{G}_v ( \vec{Q} , \overrightarrow{\nabla Q} ) \: \hat{k} \mbox{ ,}
\end{equation}

\noindent
with

\begin{equation}
\vec{\mathcal{F}} \equiv \vec{\mathcal{F}}_e - \vec{\mathcal{F}}_v \mbox{ .}
\end{equation}

\noindent
Vectors $\vec{E} \equiv \vec{E_e} - \vec{E_v}$, $\vec{F} \equiv \vec{F_e} - \vec{F_v}$ and $\vec{G} \equiv \vec{G_e} - \vec{G_v}$ are the flux vectors associated with the Cartesian coordinate triad $x$, $y$ and $z$, respectively. In the same manner, $\hat{i}$, $\hat{j}$ and $\hat{k}$ are unit vectors aligned with the same triad, respectively. The subscripts $e$ and $v$ refer to the inviscid and viscous components of each flux vector. Notice that the functional relation that exists between each vector and $\vec{Q}$ and $\overrightarrow{\nabla Q}$ is explicitly written. This is done in order to emphasize that only the viscous part of this formulation requires the evaluation of $\overrightarrow{\nabla Q}$. Lastly, $t$ is the time coordinate. The mathematical definitions of each one of these vectors are well-known in the CFD literature and, therefore, are not repeated here. The authors refer the interested reader to Ref.~\cite{HirschV1} for a complete description of the formulation.

In order to discretize Eq.~(\ref{Eq:RANS}), the FV framework is adopted. Thus, Eq.~(\ref{Eq:RANS}) is integrated over an arbitrary Eulerian domain of constant volume $\mathbb{V}$, and outer surface $S$, as follows:

\begin{equation} \label{Eq:RANSIntegralForm}
\int_\mathbb{V} \left( \frac{\partial \vec{Q}}{\partial t} \right) \: d \mathbb{V} + \int_\mathbb{V} \left(\nabla \cdot \vec{\mathcal{F}} \right) \: d\mathbb{V} = 0 \hspace*{1.0 cm} \Longrightarrow \hspace*{1.0 cm}
\frac{\partial}{\partial t} \int_\mathbb{V} \vec{Q} \: d \mathbb{V} + \oint_S \vec{\mathcal{F}} \cdot \overrightarrow{dS} = 0 \mbox{ .}
\end{equation}

\noindent
In Eq.~(\ref{Eq:RANSIntegralForm}), the Divergence Theorem, also known as the Green-Gauss Theorem, has been applied in conjunction with the general form of the Leibniz rule. Moreover, $\overrightarrow{dS} \equiv \hat{n} \: dS$, where $\hat{n}$ is the unitary normal vector that points in the outward direction of $S$.

The computational domain is assumed to be divided into multiple discrete cells of polyhedral shape, composing an unstructured grid. The discrete conserved variables vector, $\vec{Q}_i$, associated with the $i$-th cell of finite volume $\mathbb{V}_i$, is defined as

\begin{equation}
\vec{Q}_i \equiv \frac{1}{\mathbb{V}_i} \int_{\mathbb{V}_i} \vec{Q} \: d \mathbb{V} \mbox{ .}
\end{equation}

\noindent
If each cell has $n_f$ faces, then Eq.~(\ref{Eq:RANSIntegralForm}) becomes

\begin{equation} \label{Eq:FVRANS}
\mathbb{V}_i \, \frac{\partial \vec{Q}_i}{\partial t} + \sum_{k=1}^{n_f} \left( \vec{\mathcal{F}}_k \cdot \vec{S}_k \right) = 0 \hspace*{1.0 cm} \Longrightarrow \hspace*{1.0 cm}
\frac{\partial \vec{Q}_i}{\partial t} = -\frac{1}{\mathbb{V}_i} \sum_{k=1}^{n_f} \left( \vec{\mathcal{F}}_k \cdot \vec{S}_k \right) \mbox{ ,}
\end{equation}

\noindent
after applying a 1-point Gaussian quadrature rule. In the present case, this is a valid construct, since the resulting formulation is second-order accurate in space. If higher-order schemes were used instead, especially compact ones, then the surface integral might need to be numerically performed by means of higher-order quadrature rules, to which the present formulation would need to be further enhanced.

Equation (\ref{Eq:FVRANS}) is the finite volume discrete form of the RANS equations and must be true for all cells in the domain. For a mesh of constant geometry, the face area vectors, $\vec{S}_k$, are known at all times. Consequently, only two procedures are yet to be established: the reconstruction scheme used for the evaluation of the face flux vectors, $\vec{\mathcal{F}}_k$, as well as a procedure for integrating $\frac{\partial \vec{Q}_i}{\partial t}$ over time. 
Here, Roe's second-order TVD scheme, coupled with Venkatakrishnan's limiter \cite{venkatakrishnan1995}, is employed in the discretization of all inviscid fluxes \cite{bigarella2009, bigarella2012}, including the ones related to the turbulence model. The integration of the temporal derivatives is performed by using an implicit time-march scheme, as described in Refs.~\cite{bigarella2009, bigarella2012}. These schemes were chosen as part of an effort to improve the overall robustness of the solution process. Thus, the only remaining issue is to define a scheme for computing the viscous fluxes.

The calculation of the viscous components of $\overrightarrow{\mathcal{F}}_k$ requires the reconstruction of both $\vec{Q}$ and $\overrightarrow{\nabla Q}$ at the $k$-th cell face. In the present work, a standard centered approach is followed. Therefore, if $i$ and $j$ are the indexes of two adjacent cells, then:

\begin{equation}  \label{Eq:QReconstruction}
\vec{Q}_k = \frac{\vec{Q}_{k_i} + \vec{Q}_{k_j}}{2} \mbox{ ,}
\end{equation}

\noindent
in which $\vec{Q}_{k_i}$ and $\vec{Q}_{k_j}$ are the piece-wise reconstructed properties of $i$ and $j$, respectively, evaluated at the centroid of $k$ \cite{barth1989}.

Based on the same idea, $( \overrightarrow{\nabla Q} )_k$ is also reconstructed as a function of the directly adjacent cell discrete properties. The definition of this function is what sets the gradient reconstruction procedures apart from each other. In the next subsection, the three gradient reconstruction procedures considered here are briefly presented.

\subsection{Gradient Reconstruction Procedures}

\subsubsection{Weighted Green-Gauss Gradient Computation}

As previously mentioned, the evaluation of a property gradient at a cell interface usually revolves around the definition of a function with local stencil, responsible for reconstructing the gradient value at the desired location. The selection of a suitable reconstruction scheme can depend on the problem configuration, mesh geometry and even on the available computational resources. In the present work, three different gradient reconstruction procedures are considered: $A00$, $A0E$ and $AJ0$, following the naming conventions from Ref.~\cite{jalali2014}. It must be made clear that other schemes do exist \cite{jalali2014, nishikawa2010, nishikawa2011}, but only these three are analyzed here due to their simplicity and overall efficiency.

Before proceeding with a proper description of each scheme, it is important to define a method for computing the discrete cell property gradient, of which all three distinct schemes herein considered are a function of. If $A$ is a property whose discrete values, $A_i$, are known at each cell, then its gradient, $(\nabla A)_i$, can be computed as

\begin{equation}
( \overrightarrow{\nabla A} ) _i \equiv \frac{1}{\mathbb{V}_i} \int_{\mathbb{V}_i} \overrightarrow{\nabla A} \: d \mathbb{V} = \frac{1}{\mathbb{V}_i} \oint_S A \: \overrightarrow{dS} = \frac{1}{\mathbb{V}_i} \sum_{k=1}^{n_f} A_k \: \vec{S}_k \mbox{ ,}
\end{equation}

\noindent
which is referred to as the Green-Gauss approach for defining discrete cell gradients \cite{jiriBlazek2015}. The term $A_k$ can, then, be computed by using some sort of average between the adjacent known values, since it is related to the diffusive components of the original partial differential equation. Here, a volume-weighted average is used, as follows:

\begin{equation}
A_k = \frac{\mathbb{V}_i A_i + \mathbb{V}_j A_j}{\mathbb{V}_i + \mathbb{V}_j} \mbox{ .}
\end{equation}

\noindent
More robust, but more computationally expensive, schemes for computing cell-averaged property gradients are also available in the literature, such as the Linear Preserving Gradient (LPG) and the Least Squares (LS) methods \cite{cary2009}.

\subsubsection{Procedure A00}

The first gradient reconstruction procedure presented here, $A00$, is perhaps the simplest formulation possible. It consists of a simple average between the two directly adjacent cell values:

\begin{equation} \label{Eq:A00}
( \overrightarrow{\nabla A} )_k = \frac{ (\overrightarrow{\nabla A})_i + ( \overrightarrow{\nabla A})_j}{2} \equiv ( \overline{\nabla A} )_k \mbox{ .}
\end{equation}

\noindent
This reconstruction can also be improved by, instead, using a weighted average \cite{jalali2014}, without loss of computational efficiency. However, only the formulation shown in Eq.~(\ref{Eq:A00}) is considered here.

Although extremely cheap to compute, the usage of this scheme results in a stencil that effectively does not utilize information from the $i$ and $j$ cells \cite{jiriBlazek2015, weiss1999}. In turn, high-frequency errors can develop in the solution \cite{jalali2014}. To solve this problem, the formulation from Eq.~(\ref{Eq:A00}) is augmented by the introduction of extra terms that ensure dependency on cell-averaged data of the two cells that share the interface. Schemes $A0E$ and $AJ0$ are inserted in this category.

\subsubsection{Procedure A0E}

The $A0E$ scheme, also known as the edge-normal scheme, is one of the possible solutions for the previously mentioned $A00$ problem. It consists in exchanging the gradient component in the direction that connects the $i$ and $j$ cell centroids with a finite difference construct \cite{jalali2014, weiss1999}. Following Fig~\ref{Fig:cellDiagram}, the $A0E$ formulation can be written as

\begin{equation} \label{Eq:A0E}
( \overrightarrow{\nabla A} )_k = ( \overline{\nabla A} )_k + \left[ \frac{A_j - A_i}{\left| \vec{r}_{ij} \right|} - ( \overline{\nabla A} )_k \cdot \frac{\vec{r}_{ij}}{\left| \vec{r}_{ij} \right|} \right] \frac{\vec{r}_{ij}}{\left| \vec{r}_{ij} \right|} \mbox{ .}
\end{equation}

\noindent
Hence, cells $i$ and $j$ are effectively reintroduced to the stencil of $( \overrightarrow{\nabla A} )_k$.

\begin{figure}
\centering
\includegraphics[width=.7\textwidth]{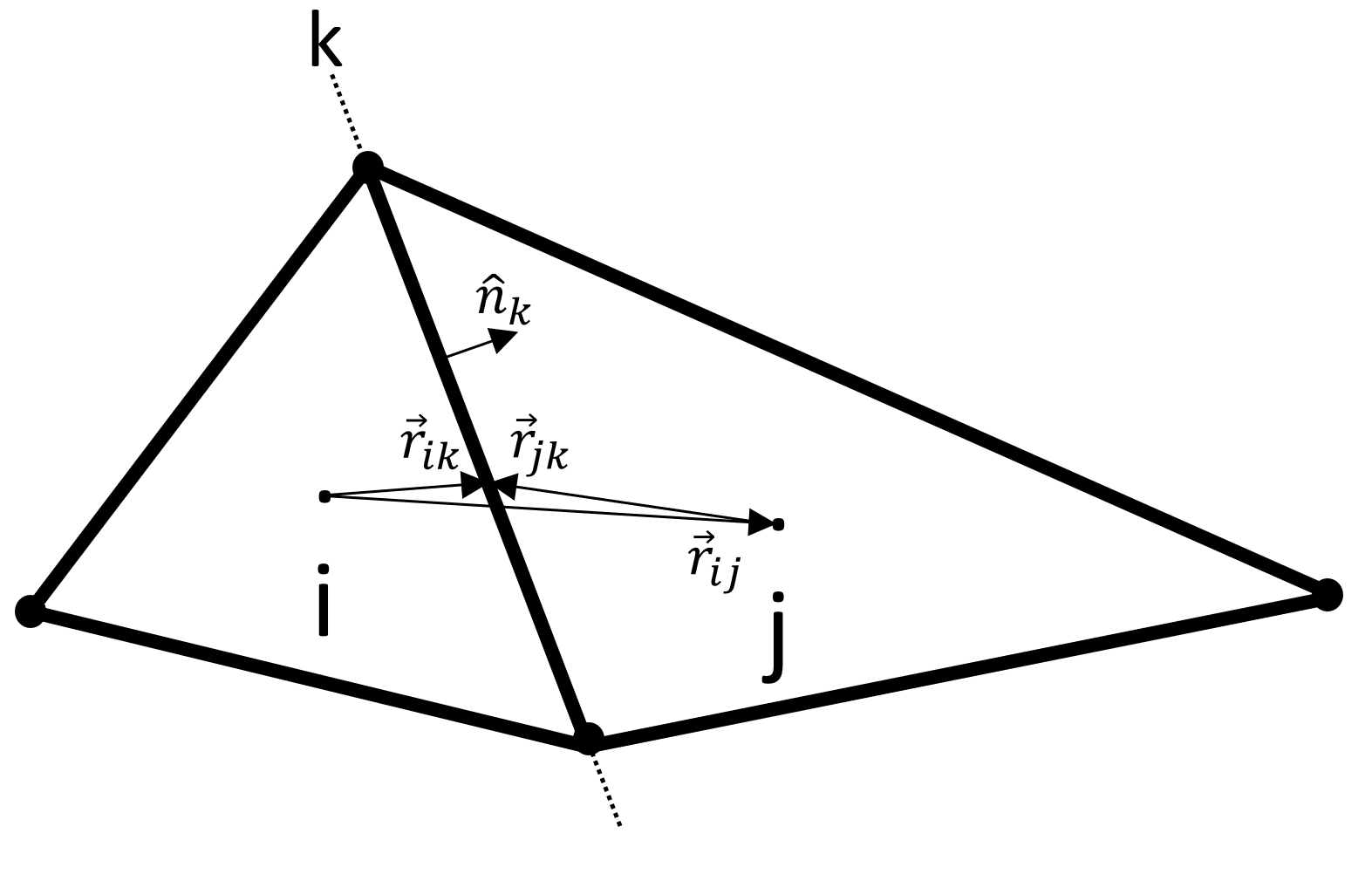}
\caption{Mesh schematic diagram picturing cells $i$ and $j$. Focus is given to the $k$-th face of the $i$-th cell. Face normal unitary vector, $\hat{n}_k$, as well as relevant distance vectors, $\vec{r}$, are also shown. The dots are used to represent the cell centroid locations.}
\label{Fig:cellDiagram}
\end{figure}

\subsubsection{Procedure AJ0}

Another approach is to use a jump term construct, $AJ0$, in which information from the discontinuous solution at the face center is introduced to the face gradient reconstruction \cite{jalali2014, nishikawa2010}. The equation, then, becomes

\begin{equation} \label{Eq:AJ0}
( \overrightarrow{\nabla A} )_k = ( \overline{\nabla A} )_k + \frac{\alpha}{\left| \vec{r}_{ij} \cdot \hat{n}_k \right|} \left( A_{k_j} - A_{k_i} \right) \hat{n}_k \mbox{ ,}
\end{equation}

\noindent
where $A_{k_i}$ and $A_{k_j}$ are the piece-wise linear reconstructed $A$ properties of cells $i$ and $j$, respectively, evaluated at the face centroid. Furthermore, $\hat{n}_k$ is the face normal unitary vector pointing outwards from the current cell. Lastly, $\vec{r}_{ik}$ is a vector that points from the centroid of cell $i$ to the centroid of face $k$, as seen in Fig.~\ref{Fig:cellDiagram}. For the $A0E$ scheme, the jump coefficient, $\alpha$,  is taken to be $\alpha = 4/3$.

It can be shown that multiple gradient reconstruction techniques can be cast into the form of Eq.~(\ref{Eq:AJ0}) \cite{nishikawa2010}. In fact, the $A0E$ scheme can be written by using Eq.~(\ref{Eq:AJ0}) with the following $\alpha$:

\begin{equation}
\alpha = \left( \hat{n}_k \cdot \hat{e}_{ij} \right) \left| \hat{n}_k \cdot \hat{e}_{ij} \right| \mbox{ ,}
\end{equation}

\noindent
where

\begin{equation}
\hat{e}_{ij} \equiv \frac{\vec{r}_{ij}}{\left| \vec{r}_{ij} \right|} \mbox{ .}
\end{equation}

\noindent
The above expression for the $A0E$ scheme is the one that is effectively implemented here.

\section{Description of Test Cases}

In this section, a brief description of each test case is presented.

\subsection{Zero-Pressure Gradient Flat Plate}

\begin{figure}
    \centering
    \includegraphics[width=\textwidth]{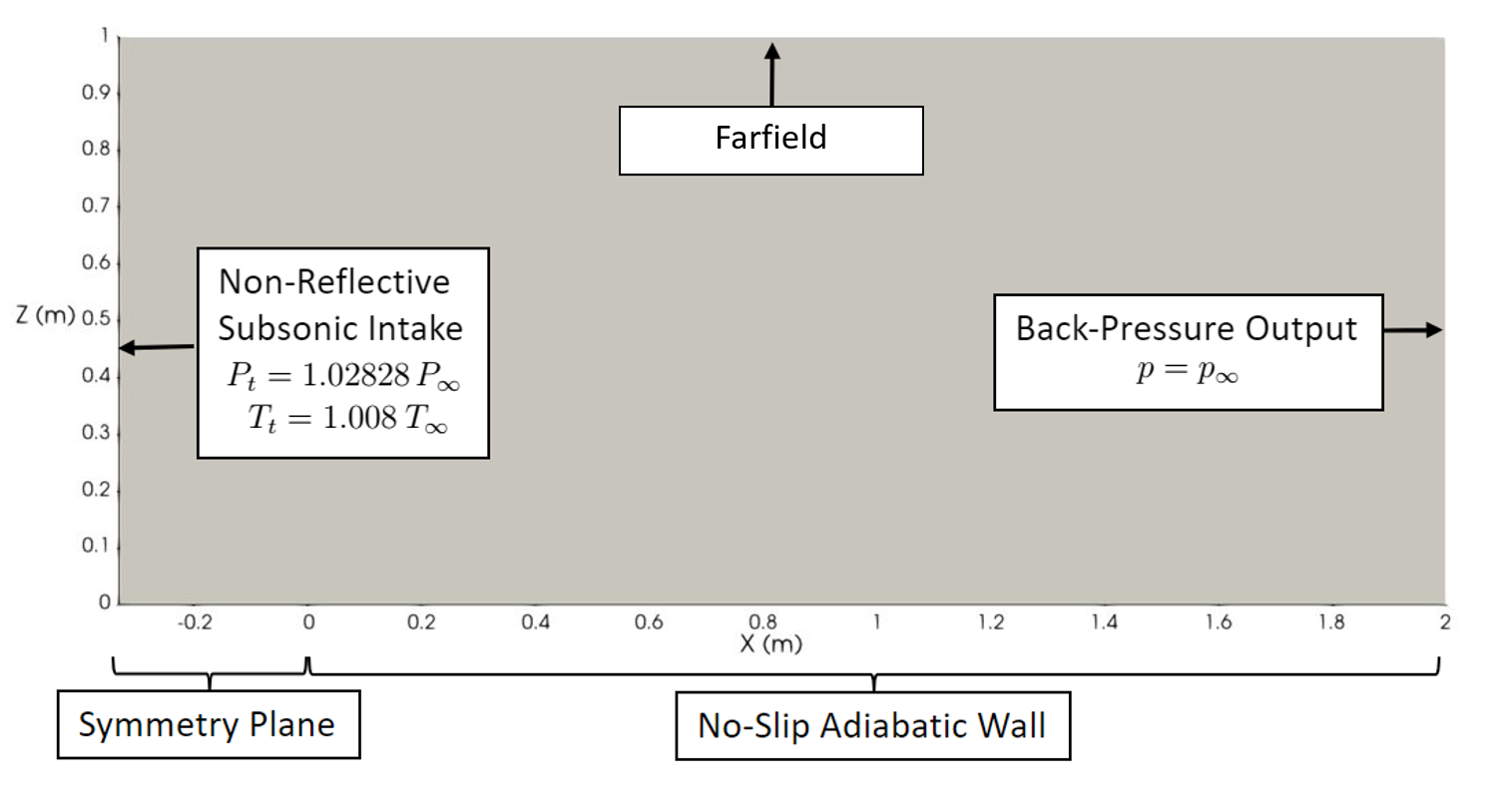}
    \caption{Boundary condition placement for the two-dimensional zero-pressure gradient flat plate case.}
    \label{Fig:flatPlateCaseDiagram}
\end{figure}

The flat plate case follows NASA Langley's Turbulence Modeling Resource (TMR) setup \cite{TMR}. Hence, it is an incompressible case solved by using a compressible fluid formulation. The problem consists of a simple rectangular domain with a length of $2.33$ m and a height of $1$ m. The first $0.33$ m of the bottom boundary is a symmetry plane. An infinitely thin flat plate, which is modeled as an adiabatic no-slip wall,  lies in the other $2$ m. 
The origin of the domain is located at the leading edge of the plate, with the $X$ axis parallel to the plate surface, pointing towards the right side of the domain. Moreover, the $Z$ axis points upwards. The top boundary is a non-reflective farfield, implemented using Riemann invariants. The freestream Mach number is set to $M_{\infty}=0.2$, at a static temperature of $T_\infty=300$ K and Reynolds number $Re_{\infty}=5 \times 10^6$, computed based on the reference length $\ell_{ref} = 1$ m. 
The right boundary is a simple back-pressure output, which is set to enforce the freestream static pressure $p_\infty = 114.47$ kPa. The left boundary is a non-reflective subsonic intake, with a total pressure of $p_t = 1.02828 \: p_\infty$, and a total temperature of $T_t = 1.008 \: T_\infty$. A diagram that illustrates the problem is shown in Fig.~\ref{Fig:flatPlateCaseDiagram}.

\begin{figure} 
    \centering
    \includegraphics[width=\textwidth]{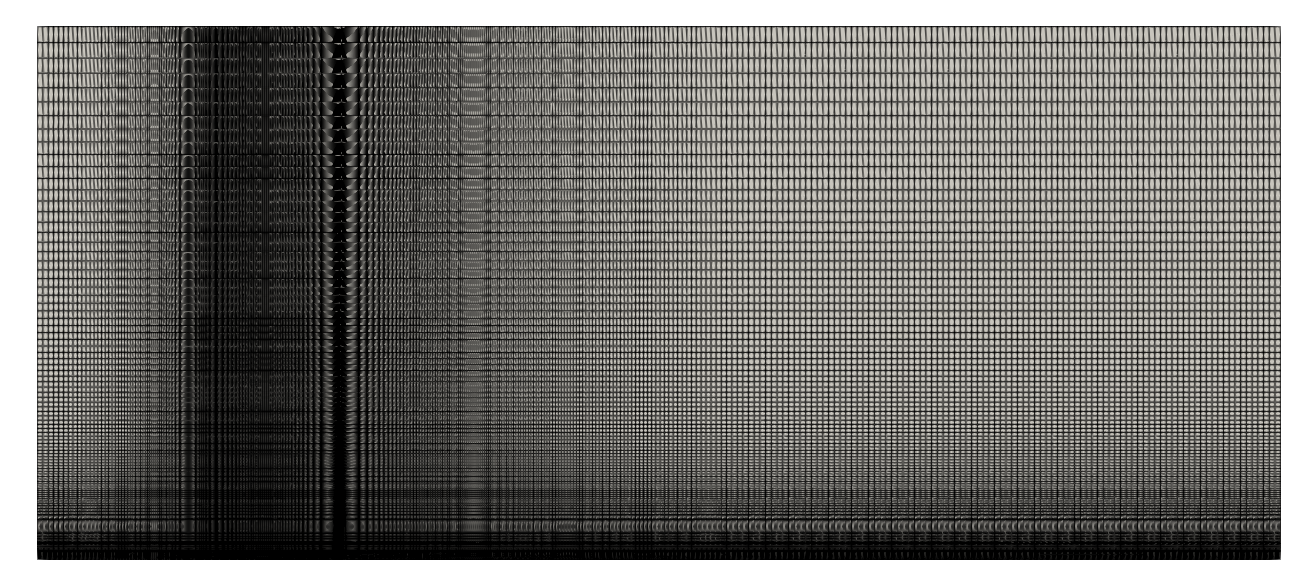}
    \caption{Mesh used in the flat plate case, containing $544$ x $384$ cells.}
    \label{Fig:flatPlateMesh}
\end{figure}

Since BRU3D is a 3-D code, the quadrilateral mesh is obtained by using hexahedral meshes with a single cell depth-wise. The mesh employed here is the finest hexahedral mesh available in the TMR website \cite{TMR}, and is composed of $544$ cells in the $X$ direction and $384$ cells in the $Y$ direction. Cells are clustered in the region near the leading edge of the flat plate, as seen in Fig.~\ref{Fig:flatPlateMesh}.

\subsection{Subsonic NACA 0012 Airfoil}

\begin{figure} [htb!]
    \centering
    \includegraphics[width=\textwidth]{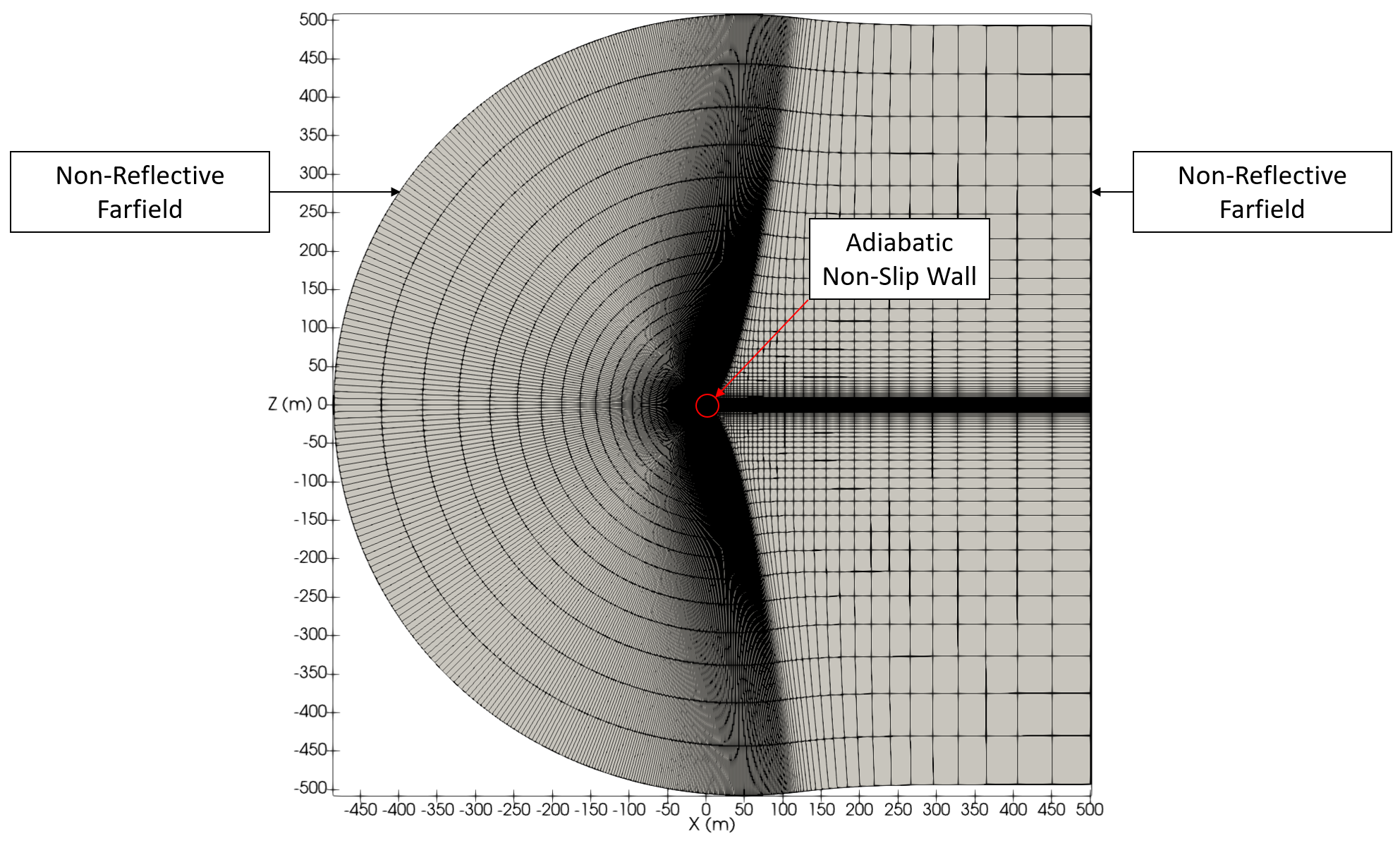}
    \caption{Boundary condition placement and overall view of the computational mesh for the subsonic NACA 0012 airfoil case.}
    \label{Fig:naca0012CaseDiagram}
\end{figure}

The NACA 0012 Airfoil case follows, once again, NASA Langley's Turbulence Modeling Resource setup \cite{TMR} for an angle of attack, $\alpha_{AoA}$, of $15$ degrees. The domain has two boundary conditions: no-slip adiabatic wall and non-reflective farfield, as shown in Fig.~\ref{Fig:naca0012CaseDiagram}. The freestream conditions, which includes the Reynolds number, $Re_\infty$, Mach number, $M_\infty$, reference chord, $c$, and reference temperature, $T_\infty$, are shown in Tab.~\ref{Tab:n0012FreestreamProperties}. The mesh used is the finest hexahedral ``C''-shaped mesh available in Ref.~\cite{TMR}. It contains $917504$ cells, mainly clustered around the airfoil surface, as illustrated in Fig.~\ref{Fig:naca0012CaseDiagram}.

\begin{table}[htb!]
\centering
\caption{Freestream conditions for the subsonic NACA0012 airfoil case.}
\begin{tabular}{c c c c c}
\toprule
\textbf{$Re_\infty$} & \textbf{$M_\infty$} & \textbf{$c$} & \textbf{$T_\infty$} & $\alpha_{AoA}$ \\ \midrule
$6 \times 10^{6}$     & $0.15$              & $1$ m        & $300$ K             & $15$ deg.      \\ \bottomrule
\end{tabular}
\label{Tab:n0012FreestreamProperties}
\end{table}

\subsection{Transonic OAT15A Airfoil}

\begin{figure}[htb!]
    \centering
    \begin{minipage}{0.48\textwidth}
        \centering
        \includegraphics[width=\textwidth]{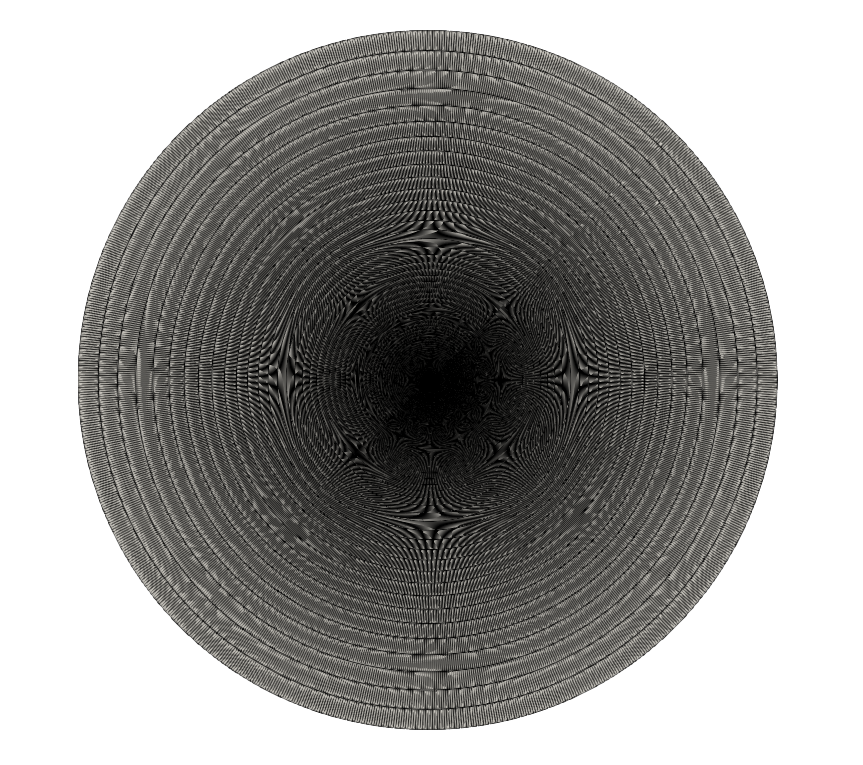}
        \caption{Overview of the mesh used in the transonic OAT15A airfoil case.}
        \label{Fig:oat15aMesh1}
    \end{minipage}\hfill
    \begin{minipage}{0.48\textwidth}
        \centering
        \includegraphics[width=\textwidth]{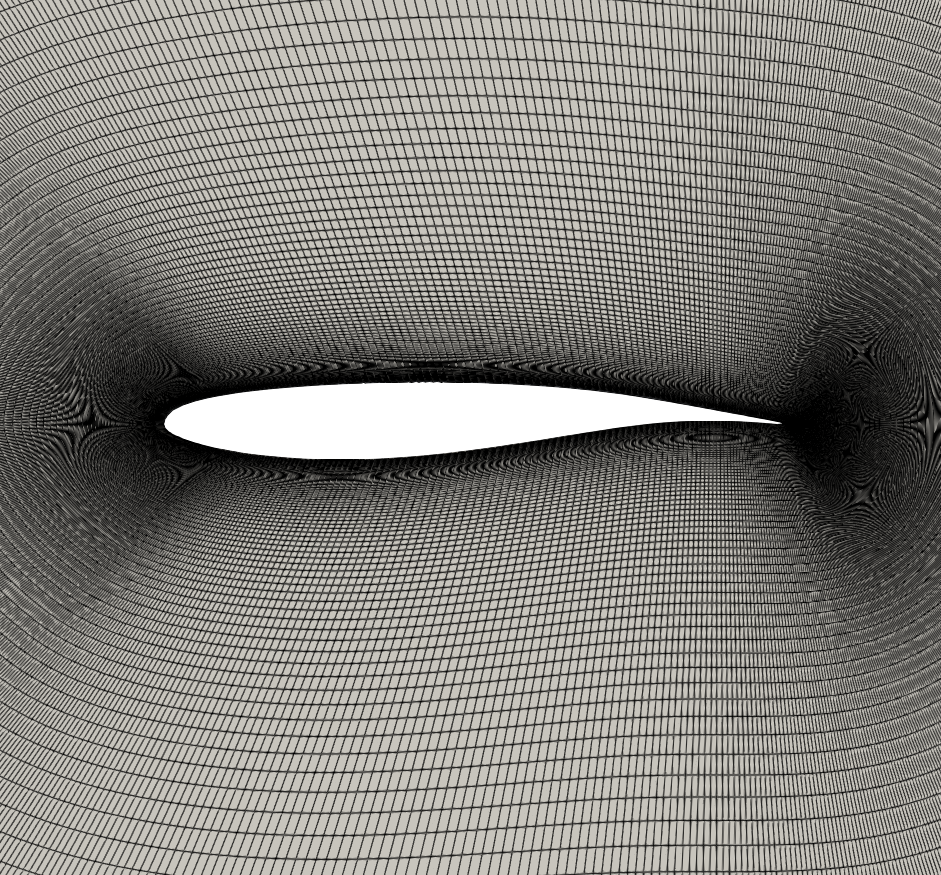}
        \caption{Zommed-in view of the mesh used in the transonic OAT15A airfoil case.}
        \label{Fig:oat15aMesh2}
    \end{minipage}
\end{figure}

The final case is the transonic OAT15A airfoil, described in Ref.~\cite{roddle1994}. Here, the boundaries are laid out in a similar manner to the previous case. That is, two continuous surfaces are employed. The innermost one is the airfoil surface, where a no-slip adiabatic boundary condition is imposed. Mereover, the outermost one is the freestream, where a non-reflective farfield is enforced. The freestream conditions are presented in Tab.~\ref{Tab:oat15aFreestreamProperties}. The mesh is constructed with 410 cells distributed along the airfoil chord. The farfield is located at 240 chords away from the airfoil surface. Cells are clustered around the airfoil surface, in such a way that $y^+ \approx 1$ at the no-slip wall, as despicted in Figs.~\ref{Fig:oat15aMesh1} and \ref{Fig:oat15aMesh2}.

\begin{table}[htb!]
\centering
\caption{Freestream conditions for the transonic OAT15A airfoil case.}
\begin{tabular}{c c c c c}
\toprule
\textbf{$Re_\infty$} & \textbf{$M_\infty$} & \textbf{$c$} & \textbf{$T_\infty$} & $\alpha_{AoA}$ \\ \midrule
$3 \times 10^{6}$     & $0.724$              & $1$ m        & $246.66$ K             & $1.15$ deg.      \\ \bottomrule
\end{tabular}
\label{Tab:oat15aFreestreamProperties}
\end{table}

\section{Results and Discussion}

In this section, the obtained results are presented, followed by a brief discussion. In all cases, the solution is considered converged when a decrease of $10$ orders of magnitude is obtained in the $L_\infty$ norm of the residue related to the continuity equation.

\subsection{Zero-Pressure Gradient Flat Plate}

\begin{figure}
\centering
\includegraphics[width=.7\textwidth]{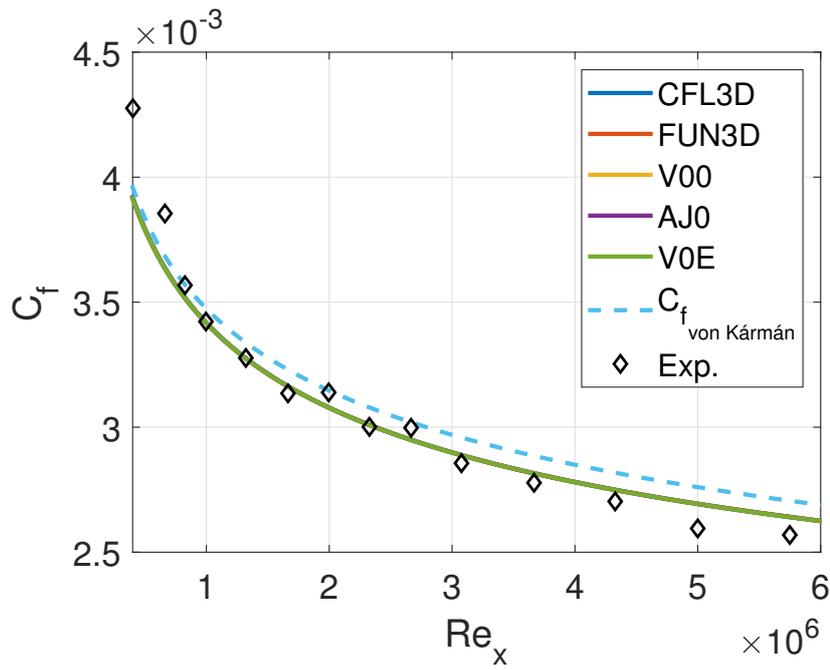}
\caption{Skin friction coefficient, $c_f$, distribution as a function of $Re_x$, along the first $1.2$ m of the flat plate. Experimental data from Ref.~\cite{coles1969} are added for comparison.}
\label{Fig:flatPlateCf}
\end{figure}

Values of skin-friction coefficient, $c_f$, plotted along the length of the flat plate are shown in Fig.~\ref{Fig:flatPlateCf}. The skin friction coefficient is defined as

\begin{equation} \label{Eq:cfDef}
c_f \equiv \frac{\tau_w}{\frac{1}{2} \rho_\infty u_\infty^2} \mbox{ ,}
\end{equation}

\noindent
in which $\tau_w$ is the fluid shear stress measured at the wall. Experimental data from Ref.~\cite{coles1969}, along with von Kármán's empirical curve \cite{white2006}, are also shown for comparison. The von Kármán empirical curve is defined as

\begin{equation} \label{Eq:cfVonKarman}
C_{f_{von Kármán}} = \frac{0.027}{\left( Re_x \right) ^\frac{1}{7}} \mbox{ .}
\end{equation}

\begin{figure}
\centering
\includegraphics[width=.7\textwidth]{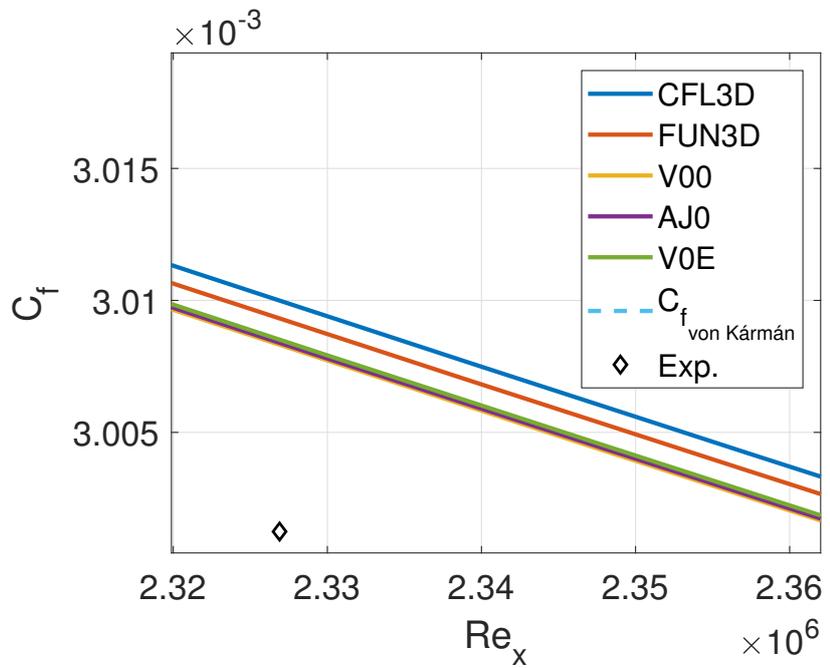}
\caption{Zoomed-in view of Fig.~\ref{Fig:flatPlateCf}, which highlights the small differences between all numerical results.}
\label{Fig:flatPlateCfZoom}
\end{figure}

Simulation data from Ref.~\cite{TMR} are also plotted. Such data was obtained with NASA's CFL3D and FUN3D codes using the Spalart-Allmaras turbulence model. In spite of the fact that slight changes can be seen between experimental data and most of the simulation data, it is clear that the results obtained by the three different gradient reconstruction schemes are virtually identical in the context of the current case setup. Furthermore, when comparing the current data with simulation results from CFL3D and FUN3D, it is also clear that they are extremely close to each other. Figure \ref{Fig:flatPlateCfZoom} shows a zommed-in view of Fig.~\ref{Fig:flatPlateCf}, which highlights the fact that the computed $c_f$ values differ from each other by a maximum of, approximately, $0.1$\%. Therefore, it is safe to say that the differences observed between the simulation data and the experimental results come from the quality of the turbulence model itself, and not from the discretization schemes used.

\subsection{Subsonic NACA 0012 Airfoil}

\begin{figure}
\centering
\includegraphics[width=.7\textwidth]{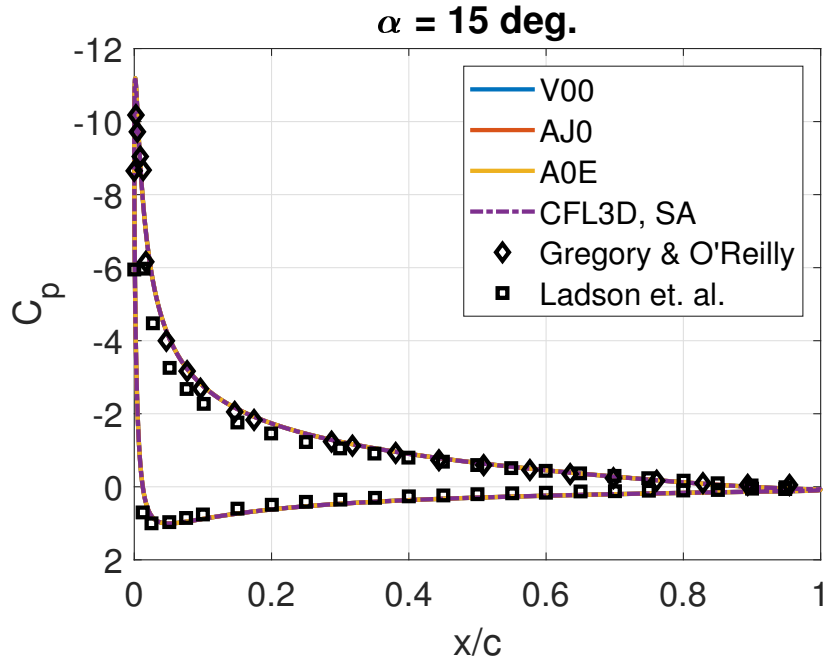}
\caption{Pressure coefficient, $c_p$, distribution along the airfoil surface for the subsonic NACA 0012 case.}
\label{Fig:naca0012Cp}
\end{figure}

Figure \ref{Fig:naca0012Cp} shows the pressure coefficient, $c_p$, plotted along the wall surface for the NACA 0012 airfoil case. Here, the pressure coefficient is computed as:

\begin{equation}
c_p \equiv \frac{\left( p - p_\infty \right)}{\frac{1}{2}\rho_\infty u_\infty^2}.
\end{equation}

\noindent
Additional simulation data from Ref.~\cite{TMR}, using the CFL3D code, as well as experimental data from Gregory \& O'Reilly \cite{mccroskey1998} and Ladson \cite{mccroskey1998} are added for comparison. As it can be seen, the same behavior previously described also repeats here. That is, no meaningful changes are captured between the schemes for the current case configuration and the present mesh topology. Furthermore, excellent agreement is observed with the experimental $c_p$ data.

\begin{figure}
\centering
\includegraphics[width=.7\textwidth]{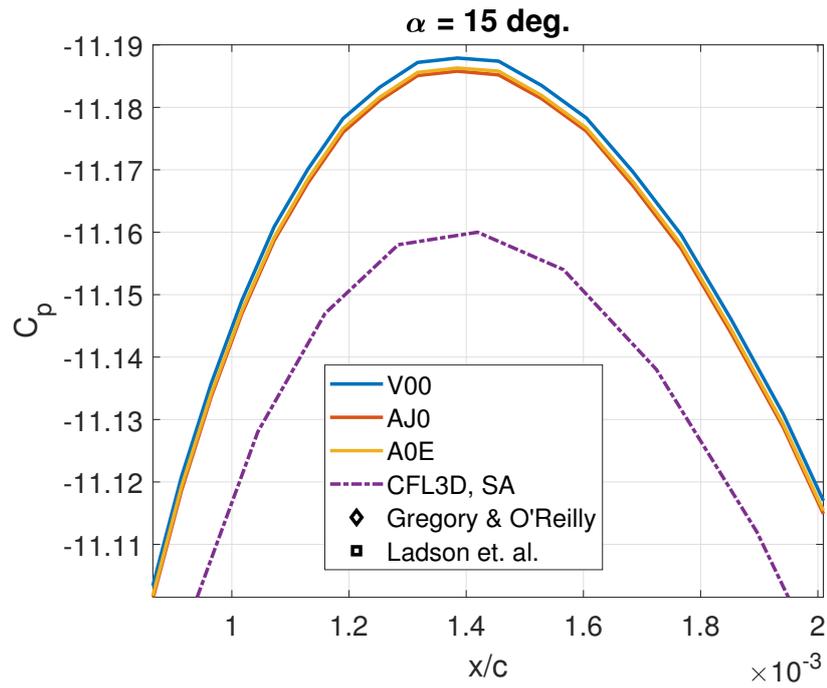}
\caption{Zoomed-in view of Fig.~\ref{Fig:naca0012Cp} in the region surrounding the suction peak.}
\label{Fig:naca0012CpZoom}
\end{figure}

In order to spot the differences between the obtained results, a zommed-in view of Fig.~\ref{Fig:naca0012Cp} is shown in Fig.~\ref{Fig:naca0012CpZoom}. Focus is given to the region surrounding the suction peak. The maximum difference between the predicted $c_p$ values is observed when the results obtained by the $V00$ scheme are compared with the ones from CFL3D. Even then, the relative difference is of only $0.27$\%, approximately.

Obtained lift and drag coefficients, $c_L$ and $c_D$, are compared in Tab.~\ref{Tab:naca0012AerCoef}. Experimental results are interpolated from Refs.~\cite{mccroskey1998} and \cite{abbott1959} and presented in the same table. It is clear that all numerical results are consistent with each other. Therefore, identical flow behavior is being captured by all numerical schemes with the SA-neg turbulence model. Any differences between the predicted coefficients and the experimental data are likely due to the turbulence model itself, and not due to the numerical discretization.

\begin{table}[htb!]
\centering
\caption{Comparison between computed aerodynamic coefficients and experimental data for the subsonic NACA 0012 airfoil case.}
\begin{tabular}{c c c}
\toprule
\textbf{$Source$} & \textbf{$c_L$} & \textbf{$c_D$}  \\ \midrule
CFL3D             & $1.5461$        & $0.02124$      \\ 
BRU3D V00         & $1.5498$        & $0.021234$     \\ 
BRU3D AJ0         & $1.5496$        & $0.021257$     \\ 
BRU3D A0E         & $1.5496$        & $0.021254$     \\ 
Exp. Gregory      & $1.5052$        & $-$            \\ 
Exp. Ladson       & $1.4993$        & $0.0180$       \\ 
Exp. Abbott       & $1.4976$        & $-$            \\ \bottomrule
\end{tabular}
\label{Tab:naca0012AerCoef}
\end{table}

\subsection{Transonic OAT15A Airfoil}

\begin{figure}
\centering
\includegraphics[width=.7\textwidth]{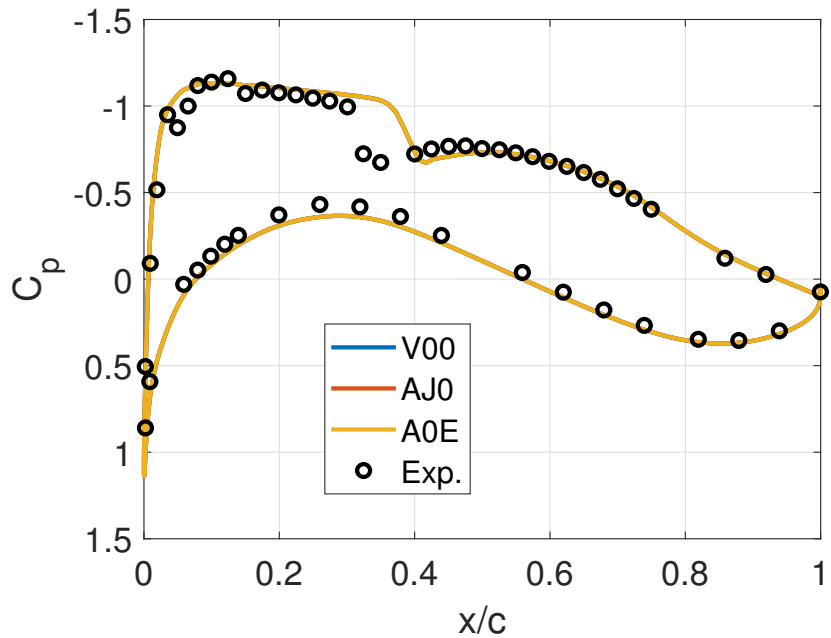}
\caption{Pressure coefficient, $c_p$, distribution along the airfoil surface for the transonic OAT15A case.}
\label{Fig:oat15aCp}
\end{figure}

\begin{figure}
\centering
\includegraphics[width=.7\textwidth]{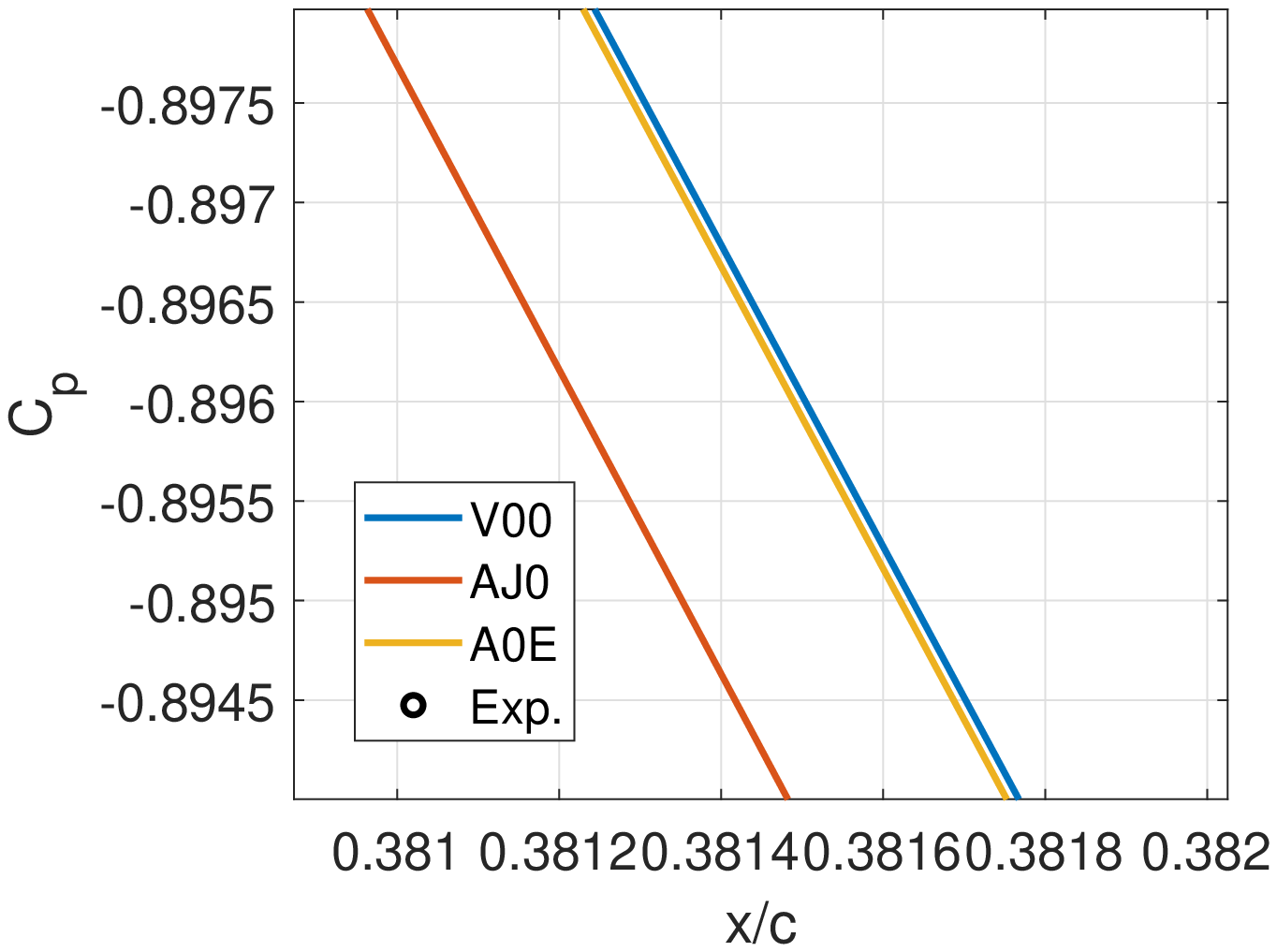}
\caption{Zoomed-in view of Fig.~\ref{Fig:oat15aCp} in the region surrounding the shock wave.}
\label{Fig:oat15aCpZoom}
\end{figure}

This is a transonic case and, therefore, shock waves are expected to develop in the numerical solution. Distribution of $c_p$ along the chord of the OAT15A airfoil is shown in Fig.~\ref{Fig:oat15aCp}, compared to experimental data from Ref.~\cite{roddle1994}. Once again, no difference is seen from the results obtained by each scheme throughout the entire length of the airfoil. This is the case even in the region surrounding the shock wave, as seen from Fig.~\ref{Fig:oat15aCpZoom}, where an extremely zoomed-in view is presented in order to visualize separate curves.

Computed aerodynamic coefficients, in the form of $c_L$ and $c_D$, are shown in Tab.~\ref{Tab:oat15aAerCoef}. Interpolated data is extracted from the plots available in Ref.~\cite{roddle1994}. There is a significant disparity between the computed coefficients and the experimental values. This is, however, a known limitation of the Spalart-Allmaras turbulence model, due to its inability to correctly solve the shock wave location for this case \cite{bigarella2009, bigarella2012}.

\begin{table}[htb!]
\centering
\caption{Comparison between computed aerodynamic coefficients and interpolated experimental data for the transonic OAT15A airfoil case.}
\begin{tabular}{c c c}
\toprule
\textbf{$Source$} & \textbf{$c_L$} & \textbf{$c_D$}   \\ \midrule
BRU3D V00         & $0.69178$        & $0.013152$     \\ 
BRU3D AJ0         & $0.69163$        & $0.013197$     \\ 
BRU3D A0E         & $0.69183$        & $0.013183$     \\ 
Interp. Exp. Data & $0.5949$         & $0.0106$       \\ \bottomrule
\end{tabular}
\label{Tab:oat15aAerCoef}
\end{table}

\section{Concluding Remarks}

Reconstruction of property gradients at cell interfaces is a small portion of the complete discretization scheme. However, it can have a profound impact on the quality of finite volume-based schemes. In the present work, essentially no differences are observed in the solutions obtained with the three different schemes. It is quite likely that such behavior is a result of the use of “well behaved” quadrilateral meshes for the three test-cases addressed here. Disparities between the computed values and experimental data are likely due to limitations of the turbulence model employed, namely, the negative Spalart-Allmaras model, and not due to discretization errors. This argument is further enhanced by comparisons with other codes that implement the same turbulence model. In these comparisons, virtually identical results are obtained. Therefore, any of the three gradient reconstruction schemes can, theoretically, be used in the simulation of cases similar to the ones investigated here.

It is important to stress, however, that these conclusions are only valid for the type of mesh considered here. Highly stretched hybrid meshes can lead to wildly different results, and the differences between each reconstruction scheme might become more obvious. This is precisely what the authors will address in future work.

\section{Acknowledgments}

The authors wish to express their gratitude to the São Paulo Research Foundation, FAPESP, which has supported the present research under the Research Grants No.\ 2021/00147-8 and No.\ 2013/07375-0. The authors also gratefully acknowledge the support for the present research provided by Conselho Nacional de Desenvolvimento Científico e Tecnológico, CNPq, under the Research Grant  No.\ 309985/2013-7. The work is further supported by the computational resources of the Center for Mathematical Sciences Applied to Industry (CeMEAI), also funded by FAPESP under the Research Grant No.\ 2013/07375-0.

\section{Contact Author Email Address}

Frederico Bolsoni Oliveira: \href{mailto:fredericobolsoni@gmail.com}{fredericobolsoni@gmail.com}, Tel.: +55 (12) 3947-6488.

João Luiz F. Azevedo: \href{mailto:joaoluiz.azevedo@gmail.com}{joaoluiz.azevedo@gmail.com}, Tel.: +55 (12) 3947-6488.

%%%%%%%%%%%%%%%% Bibliography %%%%%%%%%%%%%%%%
\bibliographystyle{AIAAbst}
\biblio{references} 
%%%%%%%%%%%%%%%%%%%%%%%%%%%%%%%%%%%%%%%%%%%%%%

\end{document}